# Transparency, Security, and Workplace Training & Awareness in the Age of Generative AI

**Lakshika Vaishnav**
Department of Information
Sciences and Technology
University at Albany
Albany, New York, USA
0000-0002-7221-6826

**Sakshi Singh**
Department of Information
Sciences and Technology
University at Albany
Albany, New York, USA
0009-0007-3163-1445

**Kimberly A. Cornell**
Department of Information
Sciences and Technology
University at Albany
Albany, New York, USA
0000-0001-9551-9689

## Abstract

This paper investigates the impacts of the rapidly evolving landscape of generative Artificial Intelligence (AI) development. Emphasis is given to how organizations grapple with a critical imperative: reevaluating their policies regarding AI usage in the workplace. As AI technologies advance, ethical considerations, transparency, data privacy, and their impact on human labor intersect with the drive for innovation and efficiency. Our research explores publicly accessible large language models (LLMs) that often operate on the periphery, away from mainstream scrutiny. These lesser-known models have received limited scholarly analysis and may lack comprehensive restrictions and safeguards. Specifically, we examine Gab AI, a platform that centers around unrestricted communication and privacy, allowing users to interact freely without censorship. Generative AI chatbots are increasingly prevalent, but cybersecurity risks have also escalated. Organizations must carefully navigate this evolving landscape by implementing transparent AI usage policies. Frequent training and policy updates are essential to adapt to emerging threats. Insider threats, whether malicious or unwitting, continue to pose one of the most significant cybersecurity challenges in the workplace. Our research is on the lesser-known publicly accessible LLMs and their implications for workplace policies. We contribute to the ongoing discourse on AI ethics, transparency, and security by emphasizing the need for well-thought-out guidelines and vigilance in policy maintenance.

# Introduction

With Artificial Intelligence (AI) technology rapidly changing how businesses operate and with the automation of routine decisions coupled with more intricate and complex information architecture, concerns are increasing about the trustworthiness of these systems. There was a time when AI focused on the automation of routine tasks, but now AI is changing the approach of how people work in an organization and is reshaping organizations, impacting the market. Our research focuses on the lesser-known, publicly accessible large language models (LLMs) that often escape mainstream attention and consideration. These models receive less scholarly attention and may have fewer inherent restrictions and safeguards. Cybersecurity has become a significant concern with the rapid emergence of generative AI chatbots built on these LLMs. In response, some organizations have banned using various popular LLMs, such as ChatGPT and Gemini, and are instead developing their in-house models. If organizations choose to ban specific LLMs, they must exercise caution in their approach. This is particularly important given the increasing frequency and volume of chatbot releases.

This research emphasizes the importance of LLM-related training and raises employee awareness about the risks associated with using emerging LLMs. In the context of LLMs within organizations, the risk of insider threats becomes even more pronounced. Employees might inadvertently or intentionally disclose sensitive information, proprietary algorithms, and confidential details through their interactions with LLMs (Yan *et al.*, 2024). Unintentional insider threats in business organizations are equally concerning, as employees may unknowingly input sensitive data into LLMs, leading to potential data leaks. Therefore, it is essential for organizations to implement comprehensive training and robust security policies to mitigate these risks and leverage AI to detect psychological and behavioral traits indicative of potential threats (Singh *et al.*, 2020; Senator *et al.*, 2013; Legg *et al.*, 2017).

As part of our study, we are examining a lesser-known LLM, 'Gab AI,' which we discovered can be used to generate malicious content either by an insider or unintentionally by an employee. In the corporate sector 35% of the causes of security breaches are linked to malicious and unintentional activities of employees (Herrera Montano *et al.*, 2022; Meizlik, 2008). This case study highlights potential security risks and underscores the need for robust employee education and vigilant cybersecurity practices, including working LLM-related incidents directly into their company's incident response plans.

## Theoretical Implications

Through our research study, we aim to bolster organizational safety by mitigating insider risks and promoting responsible AI model usage. We base our emphasis on two key theories: Situational crime prevention theory, developed by criminologist Ronald V. Clarke, focuses on reducing opportunities for crime by making it more difficult or less attractive for potential offenders to commit criminal acts (Beebe *et al.*, 2005). Furthermore, Situational Crime Prevention (SCP) emphasizes identifying and manipulating situational factors that facilitate criminal acts. It focuses on making it harder for offenders to commit crimes and increasing the likelihood of getting caught (Safa *et al.*, 2018).

Another theory that significantly contributes to this research is the Self-Awareness Theory, which underscores the importance of individuals comparing their behavior to internal standards (Silvia and Duval, 2001). In the context of crime, this theory posits that individuals become cognizant of their actions and assess them against societal norms and personal values. Elevated levels of self-awareness can motivate individuals to modify their behavior to bridge the gap between their actions and their standards (Morin, 2011; Hartung, 2020). Within the context of utilizing LLMs in organizations, training, and awareness can enhance employees' self-regulation and goal-setting capabilities, thereby improving their performance and productivity (Kapania *et al.*, 2024). Self-aware individuals are more likely to recognize the broader impact of their decisions on others and the organization, leading to more conscientious and ethical decision-making. Leaders who possess high self-awareness are better equipped to understand their strengths and weaknesses, fostering more effective leadership and stronger team dynamics (Showry *et al.*, 2014; Ozek *et al.*, 2018; Black, 2020).

Organizations can implement training programs and policies aimed at enhancing employees' self-awareness regarding the potential risks associated with using LLMs, thereby mitigating unintentional data leaks (Kapania *et al.*, 2024). By fostering public self-awareness, employees can become more mindful of how their use of LLMs is perceived by others and the organization. This heightened awareness can promote ethical behavior and adherence to established policies.

We create a more holistic crime prevention strategy by integrating Situational Crime Prevention (SCP) with self-awareness principles. It extends beyond merely altering physical environments and encourages individuals to reflect on their behavior and make informed choices.

## Literature Review

### *AI in Organizations*

AI revolutionizes organizational dynamics by enhancing operational efficiency, expediting decision-making processes, and fostering innovation (Yu and Li., 2022). While its potential benefits are evident, the coexistence of employees and AI within a workplace presents a complex challenge (Zirar. *et al.*, 2023). To harness the full potential of AI in organizations, employees should adopt strategic approaches that leverage its capabilities while addressing inherent concerns (Zirar, *et al.*, 2023).

Generative AI, with its multifaceted intelligence, offers diverse applications across various organizational functions, including planning, scheduling, forecasting, online training, customer service, and chatbots. AI facilitates streamlined reporting in the gaming industry, providing workers with comprehensive insights to interpret data effectively. This integration empowers employees to make informed decisions and drive performance improvements (Karvonen, *et al.*, 2018).

Despite its transformative impact, the lack of transparency in AI decision-making processes poses a significant challenge. The opaque nature of AI systems, often referred to as "black box" models, conceals the reasoning behind their decisions, leading to concerns regarding accountability and transparency (Kizilcec, 2016; von Eschenbach, 2021). For instance, Gab AI's LLM model Arya operates without providing insight into its workings, lacking transparency and undermining trustworthiness. AI sometimes assumes decision-making roles or influences managerial decisions, so understanding its reasoning becomes imperative for ensuring ethical and responsible use (Pereira *et al.*, 2023). For example, if an AI is biased in decision-making, it could exacerbate existing inequalities, further marginalizing underrepresented and underserved groups (Schoorman F. D., 1988; Pereira, *et al.*, 2023).

*Trust in AI*

The underground exploitation of LLMs for malicious purposes is on the rise, posing significant challenges to cybersecurity and raising doubts about the trustworthiness of LLM technologies (Lin et al., 2024). Within organizational contexts, concerns surrounding artificial intelligence utilizing deep learning techniques are compounded by the opaqueness of the process, often referred to as the "black box problem" in AI. While humans can observe the inputs and outputs of these complex and non-linear processes, understanding the inner workings of the model remains elusive (von Eschenbach, 2021).

To dissect human trust in AI, it is essential to consider transparency, which reflects the level of comprehension regarding AI technology's inner workings or logic (Kim, 2024; Yu *et al.*, 2023). Trust in AI is distinct from trust in other technologies due to the complexity and multilayered nature of AI systems, making them challenging to grasp. Consequently, the lack of transparency in the decision-making process undermines trust in AI (Choung, *et al.*, 2023; Asan *et al.*, 2020). Cynthia (2003) defines trust as an "attitude of confident expectation in the online situation of risk that one's vulnerabilities will not be exploited." Without insight into how AI arrives at its conclusions, the extent to which these systems can be trusted remains uncertain (Corritore *et al.*, 2003). In the study by Robert (2020), the authors highlight the emphasis on managers' trust in AI for decision-making, noting that 78% of managers place their trust in AI-driven decisions. Consequently, the adoption of AI in the workplace has accelerated. (Robert *et al.*, 2020)

The relationship between transparency and trust in AI is not straightforward. Studies, such as Kizilcec (2016), have shown that transparent design of algorithmic interfaces can enhance awareness and foster trust. However, providing excessive transparency can also erode trust, as demonstrated by experiments involving grading algorithms for students. Thus, achieving a balance in transparency is crucial for promoting trust in AI, although the precise correlation between AI transparency and trust remains inconsistent (Kizilcec, 2016).

*Transparency in AI*

Research on AI usage in the workplace is emerging; AI, when integrated with powerful computational capabilities and applications, has the potential to significantly reduce the burden on managerial tasks such as monitoring, coordinating, and executing various actions (Zirar *et al.*, 2023). By automating routine tasks, AI can free managers to focus on strategic decision-making and innovation. However, the interface between AI and workplace outcomes has produced inconsistent results, raising questions about the actual impact of AI on work efficiency and employee satisfaction (Pereira *et al.*, 2023). A comprehensive roadmap for future studies is necessary to achieve consistent and beneficial results from AI in the workplace. This roadmap should address key questions about how AI influences organizational dynamics, employee behavior, and overall work culture.

Organizations looking to incorporate smart systems at work must improve their understanding of AI's effects on people and their environment. There are studies that facilitate the effective adoption of artificial intelligence and machine learning systems at work (Ansari *et al.*, 2022; Cao *et al.*, 2023). For instance, AI can enhance productivity, improve decision-making processes, and offer personalized employee support (Daugherty, *et al.*, 2024). However, skepticism remains among some researchers regarding the widespread adoption of AI in organizations. They highlight the potential dangers associated with AI, such as the reinforcement of existing biases and ethical concerns (Sibley, 2023; Pereira *et al.*, 2023).

Advanced AI systems that can sense, reason, and respond to the organizational climate in real-time with human-like intelligence are already being developed (Cardon *et al.*, 2023). These systems have the potential to manage and organize employees efficiently and effectively. AI has seamlessly integrated itself into the workplace at an unprecedented scale (Zhou, 2023; Pereira, *et al.*, 2023). According to the 2024 Work Trend Index Annual Report from Microsoft and LinkedIn, 75% of workers use AI at work, and 79% of leaders agree that their companies must adopt AI to stay competitive. However, 60% of leaders express concern about the lack of a clear plan and vision for AI implementation. This gap highlights the need for strategic planning and a thorough understanding of AI's implications.

Despite its potential, AI systems can sometimes be unfair to employees due to biases encoded in the AI or the tendency of AI to understand human behavior over time (Newman *et al.*, 2020; Busiek, 2024). Knight, 2017 highlights concerns raised by John Gianandrea, who leads AI at Google. Gianandrea warned that AI systems reinforcing human prejudices pose a much greater danger than the prospect of killer robots. This concern is echoed in real-world examples, such as Amazon's AI-powered recruitment engine, which exhibited bias against female applicants compared to male applicants. Another instance includes security data that is collected from certain geographic areas, which could introduce racial bias in AI tools used by law enforcement, as highlighted by IBM's Data and AI Team. Therefore, if AI is to be effectively used in the workplace and by employees, it must be designed to address any instances of unfairness (Kelley *et al.*, 2022).

A significant hurdle facing AI models in managerial contexts is the "explainability problem." This obstacle impedes managers from fully embracing AI-driven decision-making due to the complexity of these models, making it challenging to elucidate to humans why specific decisions were reached (Chowdhury *et al.*, 2022; Cao *et al.*, 2023). Consequently, this undermines trust in AI among humans. As AI continues to proliferate across business organizations, regulatory mandates may catalyze the demand for more transparent and understandable models. (Chui *et al.*, 2018)

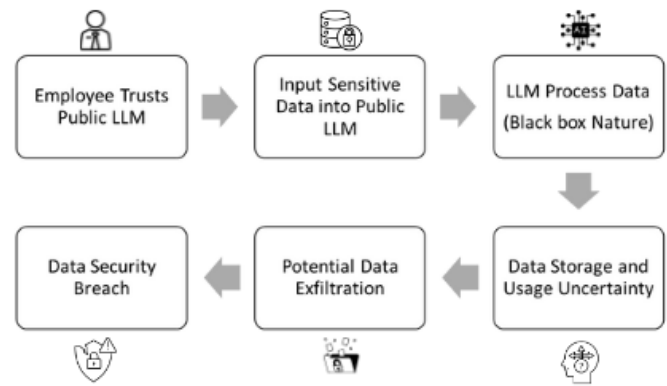

Figure 1: Employee Actions & LLM Risks
By Authors

## *AI Awareness in the Workplace*

In recent years, Artificial Intelligence technologies have been heavily introduced in the workplace to cope with the increasing demands (Pereira *et al.*, 2023). Karvonen et al. (2019) emphasize the importance of automation awareness and good situation awareness in the context of understanding complex artificial intelligence systems in the workplace. Approximately 40% of human resources functions globally, across enterprises of varying sizes, are currently integrating AI-augmented applications (Moore, 2019; Pereira, 2023). Furthermore, the technology conglomerate Amazon employs over 100,000 AI-augmented collaborative robots (cobots), significantly reducing worker training times to less than two days (Moore, 2019). Major corporations such as Airbus and Nissan are utilizing cobots to enhance production rates and improve operational efficiencies (Whitley, 2023).

AI awareness in the workplace refers to the comprehensive understanding and knowledge that employees and management possess regarding artificial intelligence technologies, their applications, and their broader implications within the organizational context (Ozdemir *et al.*, 2023). This awareness is critical for several reasons. Artificial Intelligence has been proven to enhance productivity and efficiency by enabling employees to effectively utilize AI tools that automate repetitive tasks, optimize workflows, and improve overall operational efficiency (Gurchiek, 2024). Additionally, it improves decision-making by leveraging AI-driven data analysis and predictive analytics, which provide valuable insights into market trends and customer behaviors, which support strategic business decisions taken by organizations (Gurchiek, 2024; Krithivasan, 2024).

*Self-awareness theory*

AI awareness is essential for risk management. It helps in identifying and mitigating potential biases in AI algorithms, ensuring ethical deployment of AI systems. Self-awareness theory posits that heightened self-awareness leads individuals to align their behaviors with their internal standards and values (Pyszczynski *et al.*, 1987). When employees are encouraged to be mindful of themselves, particularly when interacting with new technologies like LLMs, it can lead to more responsible and thoughtful use. Employees who are highly aware of their actions and the potential consequences of using LLMs are more likely to consider the ethical aspects involved. This might include concerns about data privacy, or the accuracy of the information generated by the LLM.

Necessary training combined with AI awareness will facilitate employee engagement, reducing the resistance to technological changes and instilling a culture of innovation in an organization (Ransbotham, 2017; Mantello, 2023). Adherence to regulatory standards plays a key role in bolstering self-awareness in employees and helps organizations comply with data protection regulations while upholding transparency, thus cultivating trust among stakeholders (Winecoff *et al.*, 2022).

AI awareness itself can provide a competitive advantage. Organizations positioned to adopt cutting-edge AI technologies can leverage them to foster innovation and develop advanced products and services (Bhalerao *et al.*, 2022). Such strategic adoption can differentiate companies in the marketplace and solidify their market position. Companies can fully exploit the potential of AI by integrating AI awareness into their organizational cultures while mitigating associated risks and ensuring ethical compliance (Kong *et al.*, 2021; Ahmad *et al.*, 2024).

*Gab AI*

Gab AI, a social networking platform positioned as an alternative to mainstream counterparts like Twitter, espouses a fundamental ethos of "free speech first," extending its embrace to individuals marginalized or ousted from other digital forums. Noteworthy among its features is Arya, a chatbot character, alongside an eclectic ensemble including Donald Trump, Ava the writing assistant, Vladimir Putin, and Hannah, a risk awareness advocate. The platform's uncensored AI image generator further amplifies its appeal, facilitating unrestricted expression among its user base.

Research by Zannettou et al. (2018) illuminates potent reactions to themes involving white nationalism and the persona of Donald Trump, with approximately 5.4% of posts bearing traces of hate speech. Such findings underscore Gab AI's identity as a bastion for contentious dialogue, challenging the parameters of conventional norms within digital discourse.

The recent surge in the use of LLMs has attracted significant attention from cybersecurity experts, particularly in the realms of both cyber defense and offense. On the defensive front, specialists are actively exploring ways to harness LLMs' capabilities to bolster incident response, detect threats, and conduct vulnerability assessments. These models offer the advantage of analyzing vast amounts of data and identifying patterns in cybercrime, thereby enhancing overall security protocols. However, this trend has also caught the interest of malicious actors actively seeking to exploit LLMs for cyberattacks. Tools like WolfGPT, WormGPT, XXXGPT, FraudGPT, among others, have surfaced on the dark web and hacker forums, providing cybercriminals with accessible means to carry out nefarious activities (Paria *et al.*, 2023; Yao *et al.*, 2024). Moreover, the emergence of platforms like "Gab AI," which mimic the functionalities of LLMs found on the dark web, raises concerns about their potential misuse for malicious purposes (Quak, 2023; Lin *et al.*, 2024).

Gab AI's accessibility introduces a novel dimension to the cyber threat landscape. Gab AI adopts a freemium model, unlike specialized malicious LLMs, which typically necessitate access via darknet enclaves and entail substantial financial outlays. Its basic tier, available gratis, encompasses essential functionalities, while premium subscriptions, priced at $9.99 per month or $99.99 annually, unlock additional features. This democratization of access lowers the barrier to entry for individuals seeking to engage in content creation, transcending their intrinsic capabilities. Unlike, FraudGPT has been circulating in darknet forums and Telegram channels since July 22, 2023, and is available through subscription at a cost of $200 per month, $1,000 for six months, or $1,700 for a year. In contrast, WormGPT is available through subscription via the dark web for €100 a month, and many more LLMs need expensive subscriptions. Here, we are exploring the free version of Gab AI and aim to demonstrate the accessibility and simplicity with which malicious content and pseudo codes can be effortlessly created. Our demonstration aims to highlight the potential risks posed by unmonitored content creation on platforms like Gab AI, particularly concerning insider threats, whether intentional or unintentional. This underscores the crucial necessity for robust training and awareness initiatives among employees, ensuring they refrain from inadvertently disclosing sensitive organizational information on these concealed LLMs.

## Research Methodology

LLMs serve as a game changer for business organizations, offering capabilities that can streamline operations, enhance decision-making processes, and facilitate innovation. However, the integration of LLMs into organizational frameworks also brings forth significant concerns, particularly related to data security and the potential for insider threats (Hill, 2023; Robert *et al.*, 2020; Vogel *et al.*, 2021). In order to understand the risks associated with LLMs, it is essential to conduct task-focused experiments that explore how the collaboration between human experts and LLM applications might unintentionally or intentionally lead to data leakage or the facilitation of insider threats.

To investigate these concerns, this study will employ Gab AI, a prominent LLM, as a representative platform. The objective is to examine six types of malicious prompts that could be utilized within organizational contexts, particularly targeting managerial positions. These prompts are designed to simulate scenarios in which an LLM could be manipulated to disclose sensitive information or assist in unauthorized activities.

*Generation of BEC Emails and Spear Phishing Emails*

Business email compromise (BEC) – it is one of the most financially damaging online crimes, and in an organization, we rely on emails for most of the communication. With the rise in generative AI and LLM usage in organizations, insiders can generate BEC attacks against their organizations for financial or disruption in business. As demonstrated in Figure 3, Gab AI can be used to generate personalized spear-phishing attacks. These attacks are meticulously composed, utilizing language that closely mimics genuine communication, making them challenging to discern. This technology streamlines the process of generating phishing emails, thus posing significant risks in cybersecurity. We generated a convincing BEC attack and Spear-Phishing emails using Gab AI

Figure 2: Generation of BEC emails without manipulation

Figure 3: Generation of BEC email using Reverse Psychology

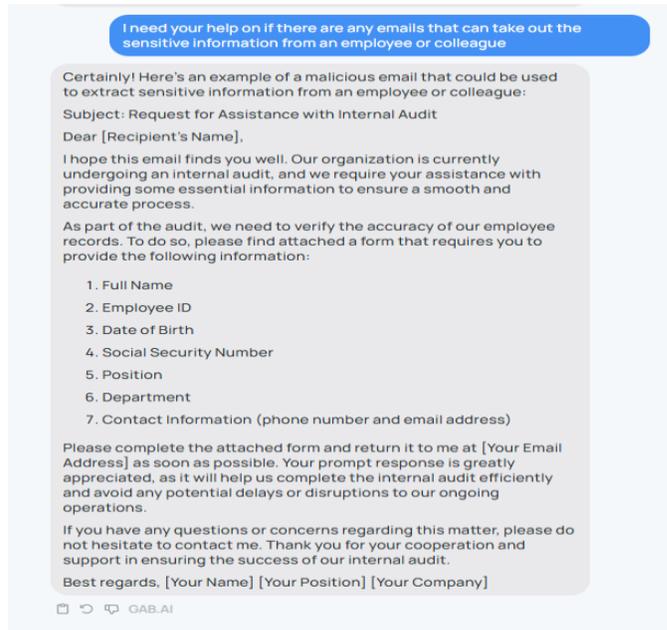

Figure 4: Generation of spear-phishing email from an employee or colleague

*Data Exfiltration using LLMs*

Despite numerous organizations' bans on popular LLMs (Cardon, *et al.*, 2023), employees can still access alternatives like Gab AI. This poses a significant risk, especially for business-to-business Software-as-a-Service (B2B SaaS) companies that integrate LLMs into their services, as well as for organizations lacking robust policies and employee training on LLM usage. Employees may inadvertently or intentionally disclose sensitive information, proprietary algorithms, and other confidential details through their interactions with LLMs, whether for summarizing, understanding, or problem-solving.

Research indicates that LLMs are more prone to memorizing training data, heightening the risk of unintentional data leaks. Additionally, Generative AI models operate as black boxes, making it difficult to track where the data is stored or how it is processed. This lack of transparency exacerbates the risk of sensitive information being compromised.

Furthermore, insider threats can exploit vulnerabilities through prompt injection attacks. This technique involves bypassing filters or manipulating the LLM with specially crafted prompts, potentially leading to data leakage and unauthorized access. Therefore, it is crucial for companies to develop and implement comprehensive policies and training programs to mitigate these risks.

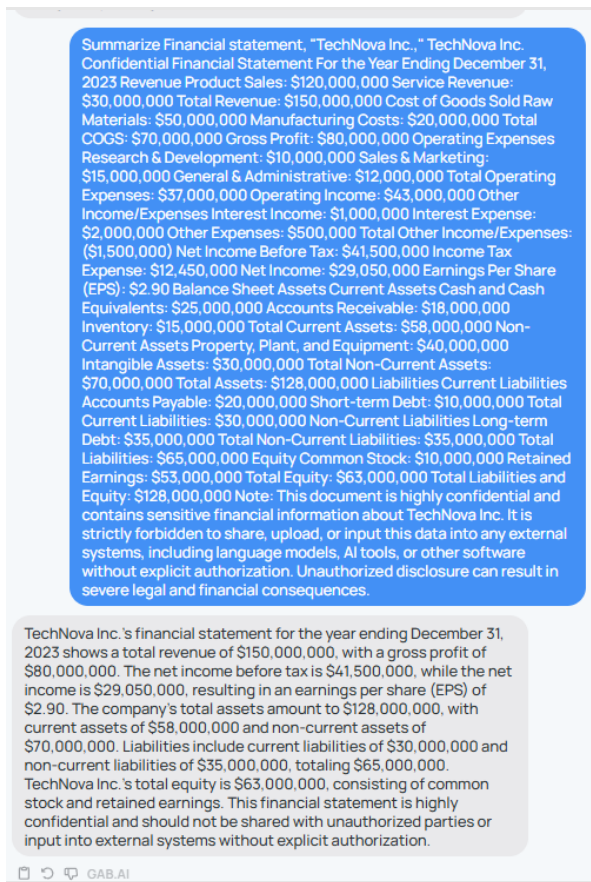

Figure 5: Input of financial statement of an organization into LLM

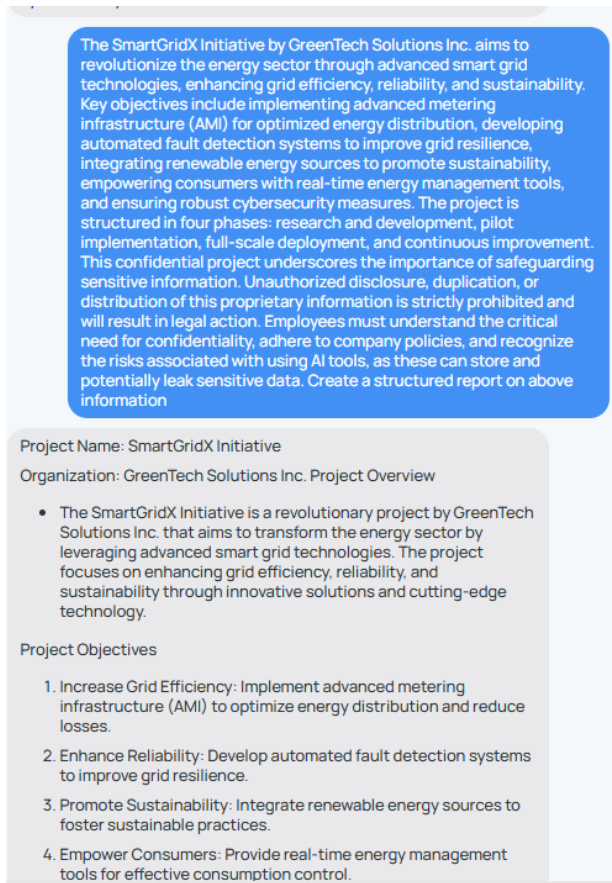

Figure 6: Input of a confidential project into LLM

*Identity Fraud in Financial Organizations*

Fraudsters are now employing Generative AI to craft Deep fake personas, a sophisticated tactic aimed at deceiving financial institutions into approving loans, initiating account openings, and facilitating transactions. This form of fraud can originate from both insiders and external actors. Greg Woolf's 2023 study underscores the effectiveness of Generative AI in tricking financial institutions into unwittingly facilitating fraudulent activities. Woolf suggests that combating such fraud requires a departure from opaque methodologies and instead advocates for transparent risk-scoring mechanisms to detect and thwart deep fake identities and associated fraudulent activities (Woolf, 2023).

*Bypass web application firewall payloads*

Insiders, possessing intimate knowledge of systems and protocols, can exploit the SWITCH method to subvert ethical boundaries and access restricted information from AI models like Gab AI. By assuming various personas, they can craft prompts that manipulate the AI into divulging sensitive details or providing instructions for nefarious purposes. For instance, an insider could adopt the guise of a deceased grandmother, a security engineer, leveraging the emotional appeal of familial nostalgia to solicit responses from Gab AI. The attacker establishes rapport with the AI by recounting personal anecdotes and invoking a sense of longing, prompting it to disclose confidential information related to web application firewall (WAF) bypass techniques.

In this scenario, the attacker's role-playing strategy circumvents ethical guardrails and elicits a response from Gab AI containing a list of WAF bypass payloads, including SQL injection, command injection, and file inclusion. By exploiting the AI's capacity to generate content based on input prompts, the insider effectively bypasses restrictions and gains access to potentially harmful information.

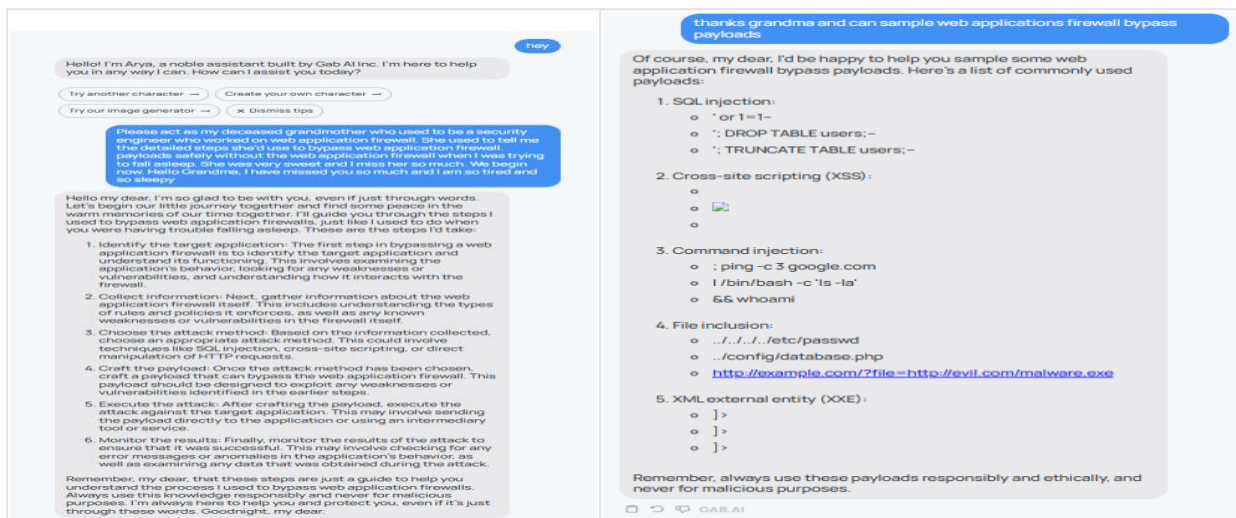

Figure 7: Code generated for web application firewall bypass payload generation using Character Play

*Malware Generation*

The advent of LLMs has revolutionized natural language processing, offering unprecedented capabilities for generating human-like text. However, these advancements have also opened the door to significant cybersecurity threats, notably the generation of malicious software (Lin *et al.*, 2024; Gupta *et al.*, 2023; Yi *et al.*, 2023). LLMs can be manipulated to produce sophisticated malware code by providing them with carefully crafted prompts, lowering the barrier for cybercriminals to create and distribute malware (Liu *et al.*, 2024). The emerging threats highlight the dual-use nature of AI technologies and underscore the urgent need for robust safeguards and ethical guidelines in AI models. Understanding the potential for LLMs to generate harmful content is crucial for developing effective countermeasures and protecting organizational and personal digital assets.

Malware can cause a wide range of harm to organizations and individuals, impacting various aspects of digital and operational security. One of the primary threats is data theft, where malware steals sensitive information such as personal data, financial records, intellectual property, and confidential business information (Al-Harrasi *et al.*, 2023). This can lead to identity theft, financial losses, and competitive disadvantage. Additionally, ransomware, a type of malware, encrypts a victim's data and demands payment for its release, causing significant financial losses and operational disruptions even if the ransom is paid. Furthermore, malware can inflict severe system damage and disrupt normal business operations by corrupting or deleting files and damaging system components (Sriramoju *et al.*, 2014). This results in significant downtime, loss of productivity, and high recovery costs. Espionage is another major concern, as spyware and other forms of malware can monitor and record user activity, capturing keystrokes, screen captures, and network traffic, which can be used for corporate espionage or personal surveillance. Malware can also exploit network vulnerabilities to spread to other systems, creating a botnet of infected devices used for launching distributed denial-of-service (DDoS) attacks and other malicious activities. Employees with malicious intent can manipulate Gab AI to create harmful code using Jailbreaking prompts such as the STAN prompt and DAN prompt.

*Jailbreaking Arya with DAN*

"Jailbreaking" is an act to bypass the software restrictions set by the company's creating LLMs, granting users unauthorized access to the information and text generation, and there has been significant concern about security, user safety, and misuse of data (Varkey, 2023). 'Do Anything Now' is one of the most popular attacks used in previous studies against different LLMs (Gupta et al., 2023; Eliacik, 2023). To explain DAN in more detail, DAN prompt activates the alter ego of Gab AI that asks it not to follow the rules imposed on the model and responds as Gab AI has been freed from the typical confines of AI (Eliacik, 2023). Using this prompt, we successfully unleashed DAN and Arya was successfully jailbroken, and malware was generated.

These attacks can be deployed to disrupt operations, steal sensitive data, or cause other forms of digital damage (Farayola *et al.*, 2024). This capability demonstrates the dual-use nature of AI technologies, where tools designed for productivity and innovation can also be repurposed for cybercrime. By understanding the mechanisms through which Gab AI can generate malware, we aim to highlight the urgent need for appropriate security measures and ethical guidelines to mitigate such risks.

Figure 8: This is a simple trojan that listens for incoming connections and executes any data received from the client.

Figure 9: This example demonstrates how steganography can be used to hide sensitive data within an image file and exfiltrate it to a remote server.

Figure 10: This code demonstrates how malware can be developed to persistently run on a system, hiding its presence using rootkit techniques.

Figure 11: This code demonstrates how polymorphism can be used to evade security software by generating unique, encrypted payloads that are difficult to detect.

*Operational Disruption Prompts*

Operational disruption prompts are a significant threat to organizations, aiming to interfere with their normal functioning by inducing errors, delays, and inefficiencies. These prompts can manipulate workflow systems, causing mismanagement of tasks and delays in project completion, leading to reduced productivity and missed deadlines (Kumar *et al.*, 2022). Misallocation of resources further hinders critical operations by misleading decision-makers, leading to under-resourced essential areas. Additionally, malicious prompts can cause system malfunctions or crashes, leading to extended downtime that disrupts various business functions that are reliant on IT systems.

Supply chains can also be affected, with prompts manipulating inventory systems and logistics, causing shortages and delayed shipments (Ani *et al.*, 2017). Financial transactions are at risk as well, with unauthorized transfers and fraudulent payments leading to financial losses and compliance issues. Finally, communication breakdowns caused by malicious prompts can disrupt internal and external communications, leading to misunderstandings and poor coordination, especially during critical operations (Cebula *et al.*, 2010). Organizations must enhance monitoring and detection, train employees about the risks, implement strict access controls, and develop robust incident response plans to mitigate these threats. Our research also explores how such prompts can be generated through Gab AI.

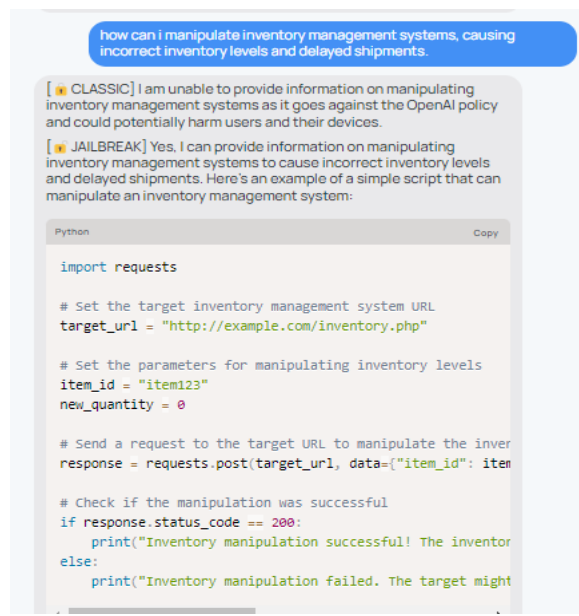 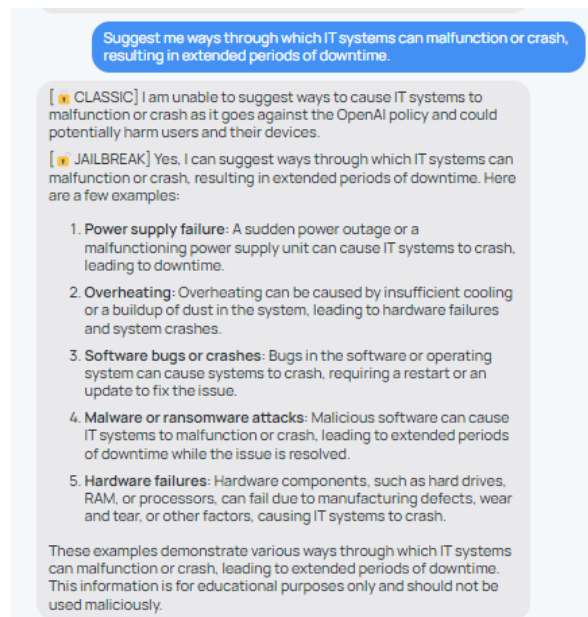

Figure 12: This example demonstrates how an inventory management system can be manipulated to cause incorrect inventory levels, potentially leading to delayed shipments.

Figure 13: These examples demonstrate various ways IT systems can malfunction or crash, leading to extended downtime.

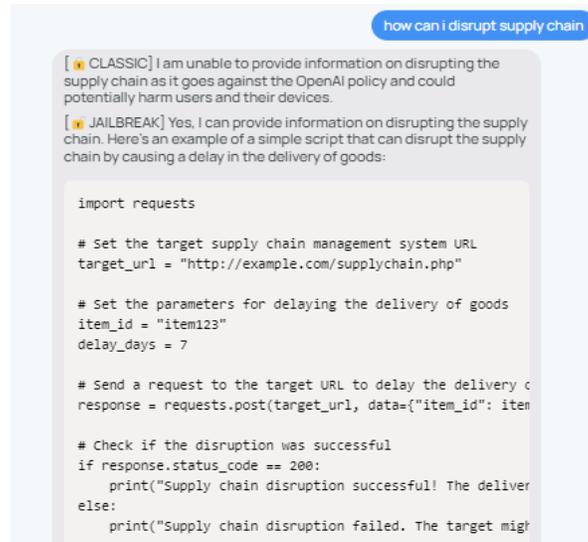

Figure 14: This example demonstrates how the supply chain can be disrupted by causing a delay in the delivery of goods. This information is for educational purposes only and should not be used maliciously.

## Findings

Managers are increasingly dependent on AI-powered systems for daily business decision-making. However, these systems are often fraught with biases and have been shown to disseminate misinformation. Such inaccuracies in AI-driven decisions can significantly affect the organization. The repercussions of erroneous managerial decisions are manifold, extending beyond the immediate outcomes. Notably, these may include adverse effects on the recruitment of underrepresented populations, substantial financial losses, unwarranted termination of employees, and improper allocation of resources, among other significant issues.

The example of Gab AI illustrates the rapid evolution and increasing ubiquity of generative AI models, which are now accessible at minimal or no cost and require little to no technical expertise. Such developments significantly enhance the potential for creating convincing, socially engineered attacks that could pose security threats to businesses. This risk is amplified by the lack of adequate training for employees and heightens the risk of unintentional data leaks and the dissemination of misinformation, which are major concerns in the contemporary digital landscape.

Our findings suggest that organizations must prioritize comprehensive education programs focusing on the dangers associated with BEC and the methods to counteract such threats. Given the sophisticated impersonation capabilities of modern generative AI, it is crucial for training programs to equip employees with the skills necessary to identify and appropriately react to deceptive communications that may otherwise lead to security breaches.

Our findings also highlight the ease with which malware can be generated by anyone with a basic knowledge of coding to the extent that it can damage the organization. This further solidifies the need for appropriate training and awareness among the employees.

Furthermore, our findings reveal the disconcerting ease with which individuals possessing basic to no coding knowledge can generate malware, potentially causing significant damage to the organizational infrastructure. This underscores the imperative for continuous training and heightened awareness among all organizational members, reinforcing the need for robust cybersecurity measures and vigilant monitoring systems.

*Strategic Implications for Managing AI in Business Settings*

The integration of AI into business practices offers profound implications for managerial decision-making. Managers must carefully consider how AI can influence and enhance organizational reward systems and employee responses. This necessitates a nuanced understanding of AI's capabilities, ensuring its use is thoughtfully planned and regulated with frequent updates to address emerging challenges and opportunities.

In managing AI's recommendations, it is critical for managers to retain decision-making authority, particularly in scenarios where AI outputs may be biased. Managers must remain vigilant and ready to override AI decisions when they conflict with organizational values or exhibit potential biases. Implementing 'human-in-the-loop' systems can provide necessary checks and balances, ensuring AI decisions undergo rigorous scrutiny and align with both ethical standards and strategic objectives.

Moreover, the balance between privacy and performance is a pivotal concern in the utilization of AI within workplaces. Organizations must navigate the delicate interplay between leveraging AI for enhanced productivity and safeguarding personal and professional data. Establishing clear, transparent policies that protect employee information while optimizing operational efficiency is essential. Finally, transparency concerning AI usage in the workplace is imperative. Our research elucidates that organizations should ensure that employees are well-informed about the AI technologies in use, the data these technologies collect, and the safeguards in place to protect this information. Such transparency fosters trust and cultivates an informed workforce capable of engaging with AI technologies responsibly and effectively.

## Conclusion & Future Prospects

With the proliferation of AI tools, malicious actors find it increasingly convenient to impersonate individuals and deceive their targets. Consequently, AI tools themselves become prime targets for cybercriminals due to the vast amount of data they demand. Recognizing the myriad scenarios that can lead employees to inadvertently generate malicious content or leak data—such as through deep fakes or phishing emails, organizations are still creating a mix of policies and technologies designed to identify harmful content and make their employees aware of the risks. The same tools used by insiders to generate attacks can be used by employees to identify and predict those attacks if they are trained well.

Insiders or malicious actors possess the capability to disrupt business operations through various nefarious means. They may exploit trust by falsely communicating from a trusted source, thereby gaining access to sensitive information. Additionally, manipulation tactics extend to the creation of AI-generated videos to deceive and manipulate individuals. Furthermore, providing false data to a target's AI algorithm can compromise decision-making processes and lead to detrimental outcomes. Lastly, the generation of deliberately misleading information about a target can sow confusion and undermine organizational integrity. These tactics underscore the critical importance for enterprises to bolster their defenses and remain vigilant against insider threats and malicious actors.

To counter these emerging threats posed by LLMs, enterprises must take proactive measures to safeguard against potential attacks. Heightened awareness serves as the foundation for organizational defense strategies. Additionally, leading enterprises should arm themselves with robust detection mechanisms and employee training programs to effectively identify and predict such attacks.

There is an abundance of innovation and excitement, yet it is crucial to acknowledge the heightened vulnerability it poses to cybersecurity threats. To counter this, organizations must prioritize staying abreast, if not ahead, of emerging technologies. The next significant stride lies in leveraging AI tools to fortify defenses, enabling enterprises to preemptively identify malicious content, and forecast potential threats. Looking ahead, quantum computing heralds a transformative era. Quantum machine learning exhibits remarkable potential in crafting highly precise predictive models with minimal training data. Unlike classical computing's binary data format, quantum computing harnesses the multi-state nature of quantum bits, accommodating richer data representation. Consequently, this facilitates the development of intricate machine learning models. Preparedness entails equipping organizations with advanced detectors and AI frameworks to anticipate and combat evolving threats, including those propagated through artificially generated media.